\documentclass{article}

\usepackage{amsmath}
\usepackage{booktabs}
\usepackage{caption}
\usepackage{graphicx, color}
\usepackage{multirow}
\usepackage{subcaption}

\usepackage[
  backend=bibtex, bibstyle=vancouver, natbib=true, sorting=none, style=numeric
]{biblatex}
\DeclareGraphicsExtensions{.pdf}

\DeclareMathOperator{\CHR}{CHR}
\newcommand{\given}[1][]{\ensuremath{#1\mid}}

\DeclareMathOperator{\OR}{OR}
\newcommand{\p}{\text{p}}
\DeclareMathOperator{\Prst}{Pr_\std}
\DeclareMathOperator{\RD}{RD}
\DeclareMathOperator{\RR}{RR}
\newcommand{\std}{\ensuremath{\text{std}}}
\newcommand{\twobytwo}{2$\times$2}

\begin{document}

\title{Rothman diagrams: the geometry of association measure modification and collapsibility}
\author{Eben Kenah}
\maketitle

\begin{abstract}
  Here, we outline how Rothman diagrams provide a geometric perspective that can help epidemiologists understand the relationships between effect measure modification (which we call association measure modification), collapsibility, and confounding.
  A Rothman diagram plots the risk of disease in the unexposed on the x-axis and the risk in the exposed on the y-axis.
  Crude and stratum-specific risks in the two exposure groups define points in the unit square.
  When there is modification of a measure of association $M$ by a covariate $C$, the stratum-specific values of $M$ differ across strata defined by $C$, so the stratum-specific points are on different contour lines of $M$.
  We show how collapsibility can be defined in terms of standardization instead of no confounding, and we show that a measure of association is collapsible if and only if all its contour lines are straight.
  We illustrate these ideas using data from a study in Newcastle, United Kingdom, where the causal effect of smoking on 20-year mortality was confounded by age.
  From this perspective, it is clear that association measure modification and collapsibility are logically independent of confounding.
  This distinction can be obscured when these concepts are taught using regression models.
\end{abstract}

\textbf{Key Messages}
\begin{itemize}
  \item Rothman diagrams, which plot risks of disease in the unexposed on the x-axis and risks of disease in the exposed on the y-axis, provide a useful geometric perspective on the relationships between confounding, association (possibly effect) measure modification, and collapsibility.
  \item When there is association measure modification on a given scale, the stratum-specific points on a Rothman diagram sit on at least two different contour lines of the measure of association.
  \item Collapsibility can be defined in terms of standardization instead of no confounding, and a measure of association is collapsible if and only if all its contour lines are straight.
  \item Rothman diagrams show that association measure modification and collapsibility are logically independent of confounding, which is difficult to see using regression model coefficients.
\end{itemize}

\section{Rothman diagrams}
\citet{rothman1975pictorial} introduced a geometric perspective on causal inference where risks of disease in the unexposed or untreated are plotted on the horizontal axis (i.e., the x-axis) and the risks of disease in the exposed or treated are plotted on the vertical axis (i.e., the y-axis).
\citet{labbe1987meta} introduced a similar plot for meta-analysis, which was used by used by~\citet{richardson2017modeling} to discuss the estimation of risk differences and risk ratios.
\citet{kenah2024rothman} showed how they can be used to illustrate the role of standardization in the control of confounding.
Here, we show they can be used to understand effect measure modification, collapsibility, and how they relate to confounding.
In recognition of the fact that changes in a measure of association across levels of a covariate $C$ can be relevant in both analytic and descriptive epidemiology, we say \emph{association measure modification} instead of effect measure modification.

As an illustration, we will use an example given by~\citet{appleton1996ignoring} from a cohort study of thyroid and heart disease among women in Newcastle, United Kingdom in 1972-1974~\cite{tunbridge1977spectrum}.
Their smoking status was measured in the original study, and their 20-year survival status was measured in a follow-up study~\cite{vanderpump1995incidence}.
Table~\ref{tab:strat} shows \twobytwo\ tables stratified by age at the time of the original survey.
Older participants were less likely to smoke but more likely to die within 20 years than younger participants, so the causal effect of smoking on 20-year mortality was confounded by age~\citep{kenah2024rothman}.
To explore effect measure modification and collapsibility, we will standardize by age group to control confounding.

\begin{table}
  \begin{subtable}{0.5\textwidth}
    \centering
    \begin{tabular}{l|rr|r}
      \toprule
      \multicolumn{4}{c}{Participants aged 18-64 years} \\
                  & Dead  & Alive   & Total \\
      \midrule
      Smoker      & 97    & 436     & 533 \\
      Nonsmoker   & 65    & 474     & 539 \\
      \midrule
      Total       & 162   & 910     & 1,072 \\
      \bottomrule
    \end{tabular}
  \end{subtable}
  \begin{subtable}{0.5\textwidth}
    \centering
    \begin{tabular}{l|rr|r}
      \toprule
      \multicolumn{4}{c}{Participants aged $\geq 65$ years} \\
                  & Dead  & Alive   & Total \\
      \midrule
      Smoker      & 42    & 7       & 49 \\
      Nonsmoker   & 165   & 28      & 193 \\
      \midrule
      Total       & 207   & 35      & 242 \\
      \bottomrule
    \end{tabular}
  \end{subtable}
  \caption{
    Age-stratified \twobytwo{} tables for smoking and 20-year mortality adapted from~\citet{appleton1996ignoring}.
    Ages are those at the time of the original survey in 1972-1974.}
  \label{tab:strat}
\end{table}

\section{Effect measure modification and contour lines}
On a Rothman diagram, every point $(x, y)$ in the unit square $[0, 1] \times [0, 1]$ represents a risk in the unexposed (the x-coordinate) and a risk in the exposed (the y-coordinate)~\cite{rothman1975pictorial, kenah2024rothman}.
We can use these risks to calculate measures of association such as the risk difference, risk ratio, odds ratio, risk ratio, or cumulative hazard ratio.
If we think of the risk difference as a function
\begin{equation}
  \RD(x, y) = y - x
\end{equation}
then the set of points with $\RD(x, y) = m$ for a given $m$ is called a \emph{contour line} or \emph{contour} of the risk difference.
Each possible value $m$ corresponds to a different contour line.
Contour lines of a measure $M$ never intersect because the point at the intersection would have at least two different values of $M$.

\begin{figure}
  \centering
  \includegraphics[width = \textwidth]{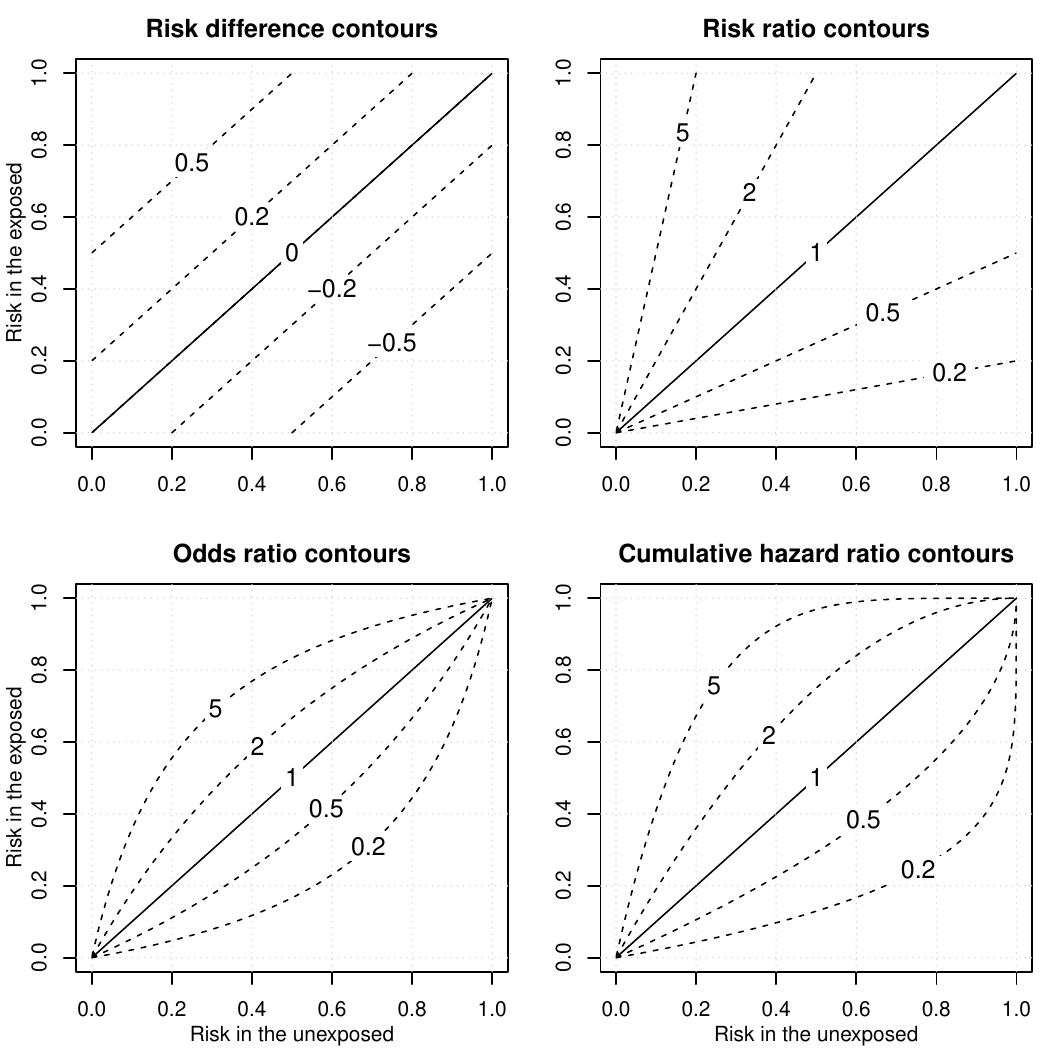}
  \caption{
    Contour lines for the risk difference, risk ratio, odds ratio, and cumulative hazard ratio.
    The null line is solid.
    All other contours lines are dashed.
    Each contour line is labeled with the corresponding value of the measure of association.
  }
  \label{fig:contours}
\end{figure}

Figure~\ref{fig:contours} shows contour lines for the risk difference as well as the risk ratio
\begin{equation}
  \RR(x, y) = \frac{y}{x}
\end{equation}
(when $x > 0$), the odds ratio
\begin{equation}
  \OR(x, y) = \frac{y / (1 - y)}{x / (1 - x)},
\end{equation}
(when $0 < x < 1$ and $y < 1$), and the cumulative hazard ratio
\begin{equation}
  \CHR(x, y) = \frac{\ln(1 - y)}{\ln(1 - x)}
\end{equation}
(when $x < 1$ and $y < 1$).
The cumulative hazard ratio equals the hazard ratio when the hazard ratio is constant.
The null line where $x = y$ is a contour line for all measures of association, corresponding to $\RD(x, y) = 0$ and
\begin{equation}
  \OR(x, y) = \RR(x, y) = \CHR(x, y) = 1.
\end{equation}
Figure~\ref{fig:contours} is completely generic, not tied to the results of any particular study.
All contours of the risk difference and risk ratio are straight, but all contours of the odds ratio and cumulative hazard ratio (except the null line) are curved.

A covariate $C$ is an association measure modifier on the scale of a measure of association $M$ if the strata defined by $C$ have at least two different values of $M$.
On a Rothman diagram, the stratum-specific points defined by levels of an association measure modifier $C$ are on at least two different contours of $M$.
If we randomly choose two or more points in the unit square, it would be an astonishing coincidence if they were all on exactly the same contour of a given $M$.
Thus, effect measure modification should be seen as normal, not pathological or unusual~\cite{greenland1987interpretation}.
It would be less surprising to find a contour line that passes close to all points, in which case a common estimate of $M$ could be a reasonable approximation.

Table~\ref{tab:modification} shows stratum-specific measures of association estimated using binomial GLMs with an interaction term to allow the estimated association between smoking and 20-year mortality to vary by age group.
The model uses an identity link to estimate the risk difference, a log link to estimate the risk ratio, and logit link (i.e., logistic regression) to estimate the odds ratio, and a complementary log-log link to estimate the cumulative hazard ratio.
For each model, the table shows the likelihood ratio p-value for the interaction term as well as the common measure of association estimated when the interaction term coefficient is set to zero.
According to the point estimates alone, there is effect measure modification on all four scales, where smoking has a harmful effect in the 18-64 year age group and a near-null effect in the 65+ year age group.
However, assuming a constant risk difference or odds ratio approximates the data well (both $\p \geq 0.3$).
There is strong evidence of effect measure modification on the risk ratio scale ($\p = 0.01$) and weaker evidence of effect measure modification on the cumulative hazard ratio scale ($\p = 0.09$).

\begin{table}
  \centering
  \begin{tabular}{lrrrrc}
    \toprule
    Measure of                & \multicolumn{3}{c}{Stratum-specific} &  & 95\% confidence\\
    association               & 18-64   & 65+     & p-value  & Common   & interval \\
    \midrule
    Risk difference           & 0.061   & 0.002   & 0.300     & 0.052   & (0.013, 0.091) \\
    Odds ratio                & 1.622   & 1.018   & 0.353     & 1.537   & (1.119, 2.125) \\
    Risk ratio                & 1.509   & 1.003   & 0.010     & 1.062   & (0.952, 1.166) \\
    Cumulative hazard ratio   & 1.563   & 1.008   & 0.085     & 1.316   & (1.034, 1.676) \\
    \bottomrule
  \end{tabular}
  \caption{
    Stratum-specific measures of association for smoking and 20-year mortality with likelihood ratio p-values for the interaction term and common estimates with likelihood ratio 95\% confidence intervals.
  }
  \label{tab:modification}
\end{table}

\begin{figure}
  \includegraphics[width = \textwidth]{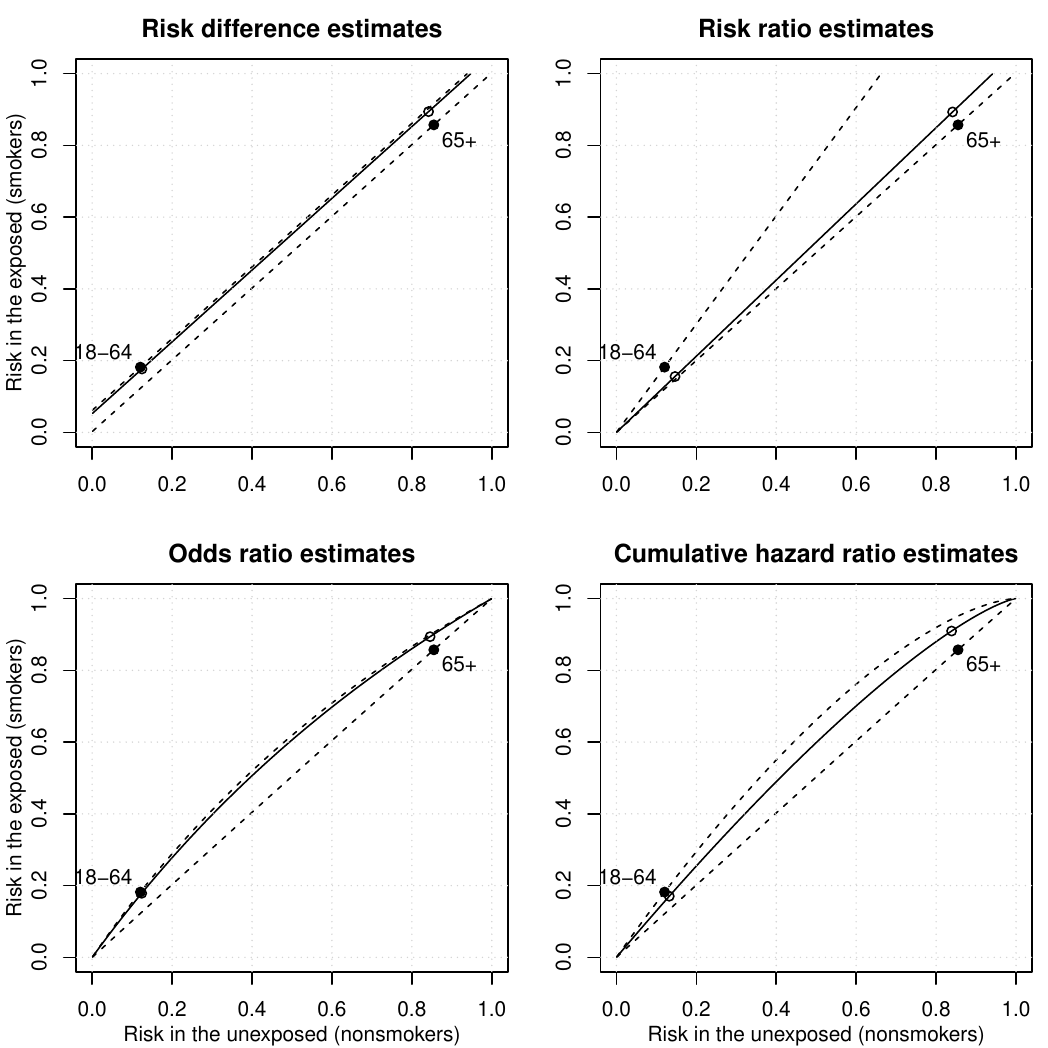}
  \caption{
    Contour lines containing the stratum-specific measures of association (dashed) and common measures of association (solid) on the risk difference, risk ratio, odds ratio, and cumulative hazard ratio scales.
    Each Rothman diagram shows the stratum-specific points from the data (filled circles) and the estimated stratum-specific points with no effect measure modification (open circles).
  }
  \label{fig:modification}
\end{figure}

The Rothman diagrams in Figure~\ref{fig:modification} show the stratum-specific points for each age group and the contour lines for the stratum-specific measures of association from Table~\ref{tab:modification}.
Because the models with an interaction term are saturated, the contours for the estimated measures of association pass through these stratum-specific points.
Each diagram also shows the estimated stratum-specific points and the contour line for the common measure of association from the model with no interaction.
Because the 18-64 year age group has 1,072 individuals while the 65+ year age group has only 242, the younger age group has a larger impact on the likelihood for all models.
On the risk difference and odds ratio scales (where there is almost no evidence of effect measure modification), the estimated stratum-specific points from the model with no interaction are close to the actual stratum-specific points---especially for the 18-64 year age group.
On the risk ratio and hazard ratio scales (where there is some evidence of effect measure modification), the estimated stratum-specific points from the model with no interaction are farther from the stratum-specific points---especially for the risk ratio in the 18-64 year age group.
These differences partly explain the different p-values in Table~\ref{tab:modification}.

\begin{figure}
  \includegraphics[width = \textwidth]{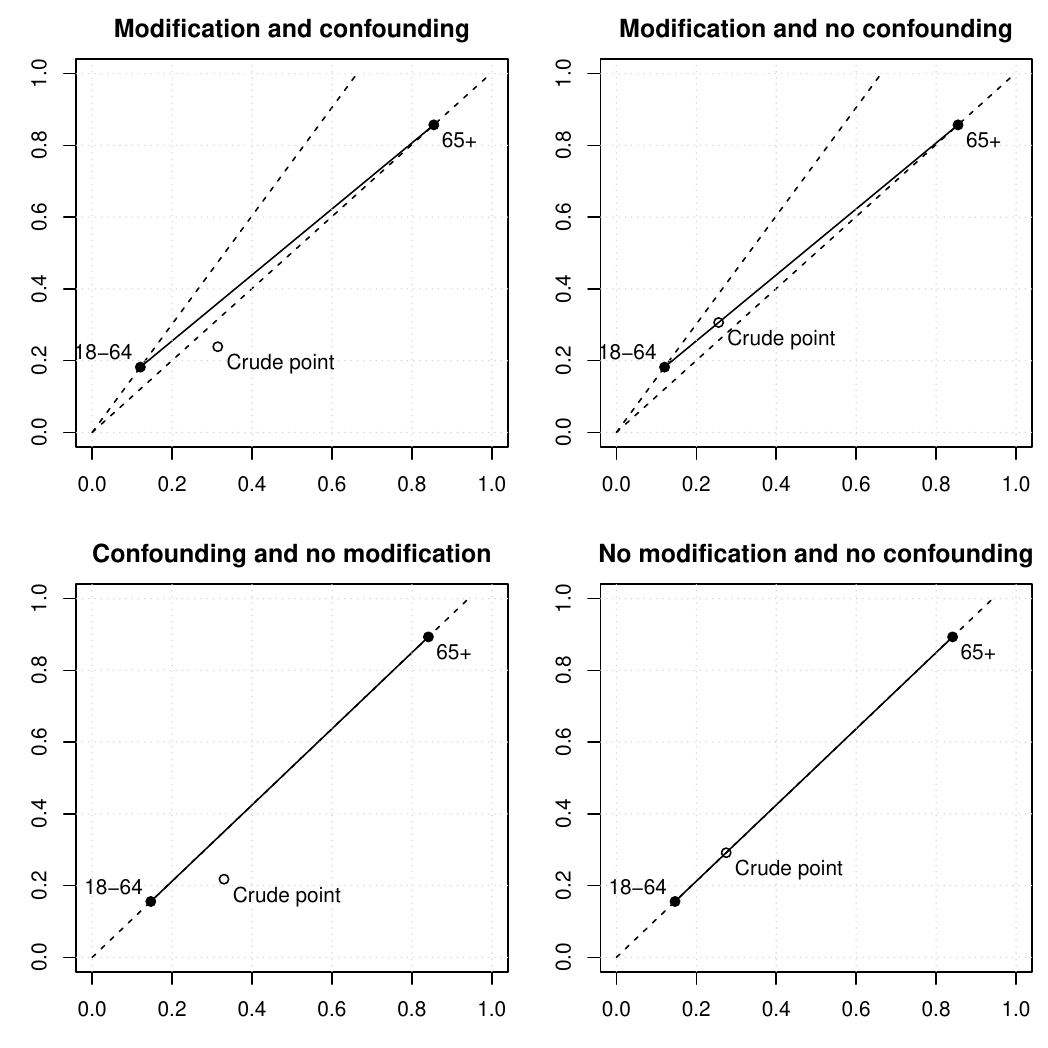}
  \caption{
    The logical independence of confounding (whether the crude point is on the standardized segment) and association measure modification (whether the stratum-specific points are on the same contour line) on the risk ratio scale for a binary covariate $C$.
    The solid line is the standardized segment, and the dashed lines are the risk ratio contour(s) for the stratum-specific points.
    The stratum-specific points with and without association measure modification are the same as in the upper right panel of Figure~\ref{fig:modification}.
    Marginal risks for smokers and nonsmokers are calculated using the exposure-specific age distributions (for confounding) or the marginal age distribution (for no confounding).
  }
  \label{fig:modconf}
\end{figure}

From a geometric point of view, it is clear that confounding and association measure modification are logically independent.
In the simple case of a binary covariate, Figure~\ref{fig:modconf} shows all four possible combinations of confounding and association measure modification on the risk ratio scale.
With a binary covariate $C$, there is confounding when the crude point (based on the marginal risks in the exposed and unexposed) is off the standardized segment, which is the line segment connecting the stratum-specific points~\cite{kenah2024rothman}.
On the scale of a measure $M$, there is association measure modification when the stratum-specific points have different values, which puts them on different contour lines of $M$.
A similar array of plots could be made for any of the other measures of association, and this logical independence extends to any covariate on any scale:
Confounding is about the relationship between the stratum-specific points and the crude point, and its occurrence does not depend on the chosen measure of association.
Association measure modification is about the relationship between the stratum-specific points and the contour lines of the chosen measure of association, and its occurrence can depend on this choice.

\section{Collapsibility and contour lines}
If $C$ is a risk factor for disease but not a confounder, a measure of association $M$ is said to be (strictly) collapsible if $M = m$ in the marginal \twobytwo\ table whenever $M = m$ in all stratum-specific \twobytwo\ tables defined by $C$~\cite{greenland1996absence, greenland1999confounding, greenland2011adjustments}.
Thus, the marginal and stratum-specific values of a strictly collapsible measure $M$ are equal whenever $C$ is not a confounder and not an association measure modifier on the scale of $M$.
The risk ratio and risk difference are collapsible but the odds ratio is not~\cite{samuels1981matching, miettinen1981confounding}.
The hazard ratio is noncollapsible~\cite{greenland1996absence}, which implies that the cumulative hazard ratio is also noncollapsible.

Let $X$ be a binary exposure or treatment, $D$ be a binary disease outcome, and $C$ be a binary covariate that is not causally affected by $X$.
In our example, $X$ is smoking status at the time of the original survey, $D$ is death within 20 years, and $C$ is age group.
When $C$ is not causally affected by $X$, the standardized risk of disease given $X = x$ and a standard distribution of $C$ is 
  \begin{equation}
  \Prst(D = 1 \given X = x)
  = \sum_{c} \Pr(D = 1 \given X = x, C = c) \Prst(C = c)
  \label{eq:prstd}
\end{equation}
where the sum is taken over all possible values of $C$.
When we calculate a measure of association using
\begin{equation}
  \begin{aligned}
    x &= \Prst(D = 1 \given X = 0), \\
    y &= \Prst(D = 1 \given X = 1),
  \end{aligned}
\end{equation}
there is no confounding by $C$ because both exposure groups have the same distribution of $C$.
The corresponding point on the Rothman diagram is a called a \emph{standardized point}.
The standardized point is always in the convex hull of the stratum-specific points, which is the region that would be enclosed by a rubber band stretched around all of these points~\citep{rothman1975pictorial, kenah2024rothman}.
This region is called the \emph{standardized hull}.
When $C$ is binary, the standardized hull is the line segment connecting the stratum-specific points, which is called the \emph{standardized segment}.

Standardization can be used to simplify the definition of collapsibility:
A measure $M$ is collapsible if and only if $M = m$ at all standardized points whenever $M = m$ at all stratum-specific points.
Because standardization removes confounding by $C$, this is equivalent to the definition that assumes no confounding by $C$.
Definining collapsibility using standardization makes it clear that collapsibility is a property of the measure of association that is independent of the stratification variable or variables.

Having redefined collapsibility in terms of standardization, we can use the geometric relationship between standardization and the convex hull of the stratum-specific points to give a geometric definition of collapsibility:
A measure $M$ is collapsible if and only if all its contour lines are straight on a Rothman diagram.
In Figure~\ref{fig:contours}, the collapsible risk difference and risk ratio have straight contour lines but the noncollapsible odds ratio and cumulative hazard ratio have curved contours away from the null.
Because the null line is a straight contour for each measure of association, they are all collapsible under the null hypothesis.

\begin{figure}
  \centering
  \includegraphics[width = \textwidth]{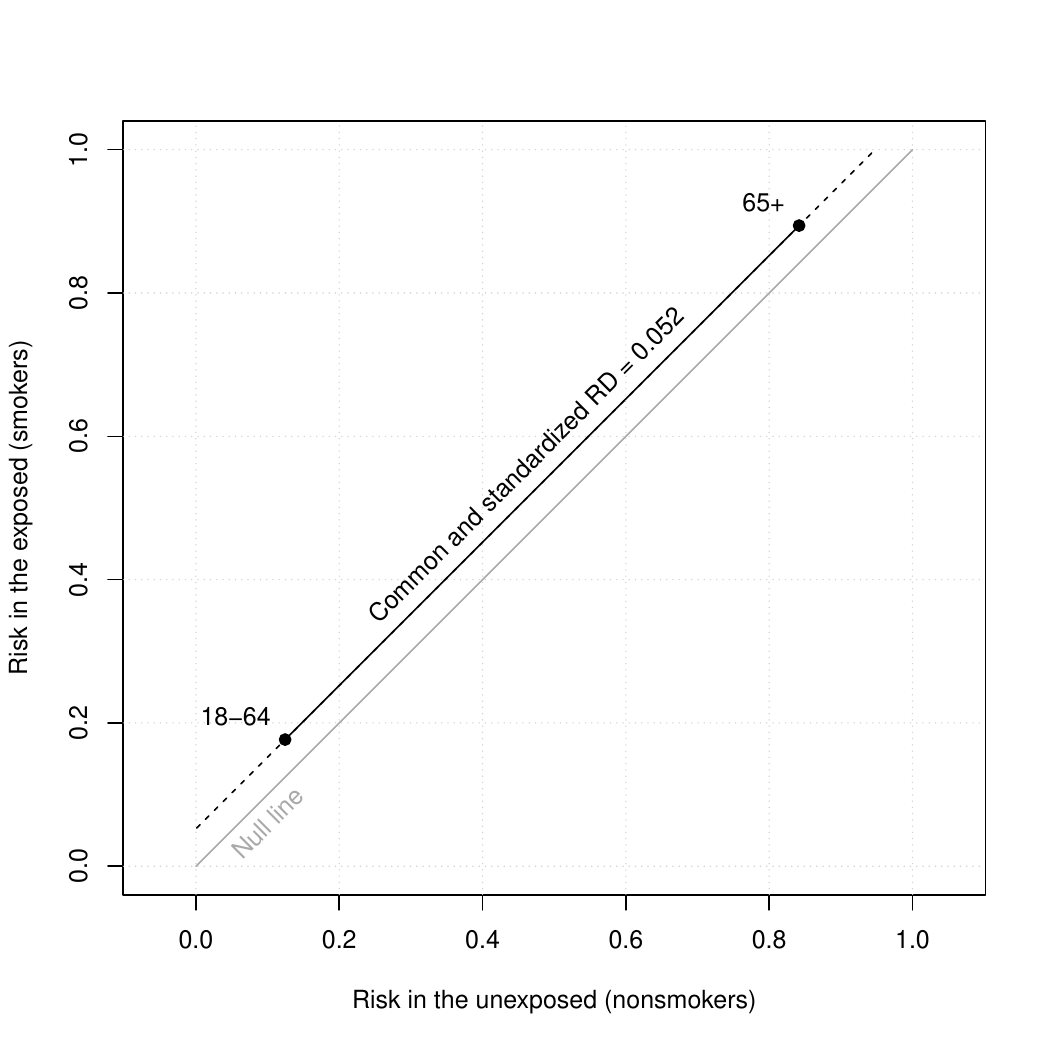}
  \caption{
    Collapsibility along a straight contour line for the common risk difference from Table~\ref{tab:modification}.
    The standardized segment connects the estimated stratum-specific points (filled circles), and it is contained entirely within the same risk difference contour.
    Thus, all standardized points have the same risk difference.
    This pattern is the same for any number of strata.
  }
  \label{fig:collapsible}
\end{figure}

For simplicity, consider the case where $C$ is binary, so the standardized hull of the stratum-specific points is just the line segment connecting them.
If all contours of $M$ are straight and both stratum-specific points on the contour line where $M = m$, then the standardized hull is a line segment along this contour, so $M = m$ at all standardized points.
Since this is true for any $m$, the measure $M$ is collapsible.
This situation is illustrated in Figure~\ref{fig:collapsible}, where all standardized risk differences equal the common stratum-specific risk difference of $0.052$ from Table~\ref{tab:modification}.
When there are more than two age strata and no effect measure modification, a collapsible measure of association (or a noncollapsible measure under the null) has more than two points along the same straight contour line but the pattern is the same.

\begin{figure}
  \centering
  \includegraphics[width = \textwidth]{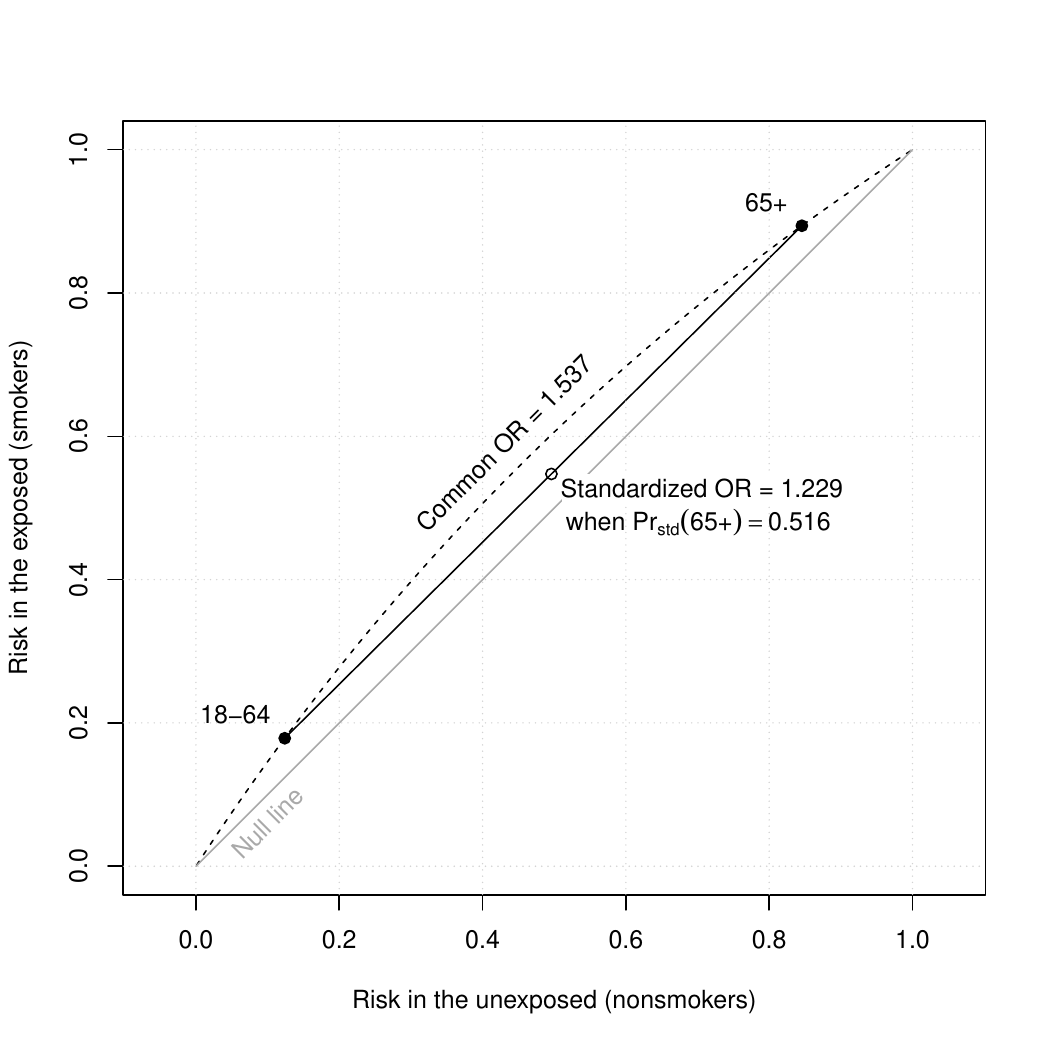}
  \caption{
    Noncollapsibility along a curved contour line for the common stratum-specific odds ratio from Table~\ref{tab:modification}.
    Except for its endpoints at the estimated stratum-specific points (filled circles), the standardized segment is between the null line and the stratum-specific odds ratio contour.
    Thus, almost all standardized points have an odds ratio closer to the null.
  }
  \label{fig:noncollapsible}
\end{figure}

If any contour $M = m$ is curved, the line segment connecting any two points $p_0$ and $p_1$ on this contour contains points where $M \neq m$, so $M$ is not collapsible.
This situation is illustrated in Figure~\ref{fig:noncollapsible}, where almost all standardized odds ratios are closer to the null than the common stratum-specific odds ratio of $1.537$ from Table~\ref{tab:modification}.
The minimum standardized odds ratio is $1.229$, which occurs in a standard population that is approximately $48.4\%$ aged 18-64 years and $51.6\%$ aged $\geq 65$ years.
The standardized odds ratio equals the common stratum-specific odds ratio only if the standard population consists entirely of one age group, in which case the standardized point will be one of the endpoints of the standardized segment.

\begin{figure}
  \centering
  \includegraphics[width = \textwidth]{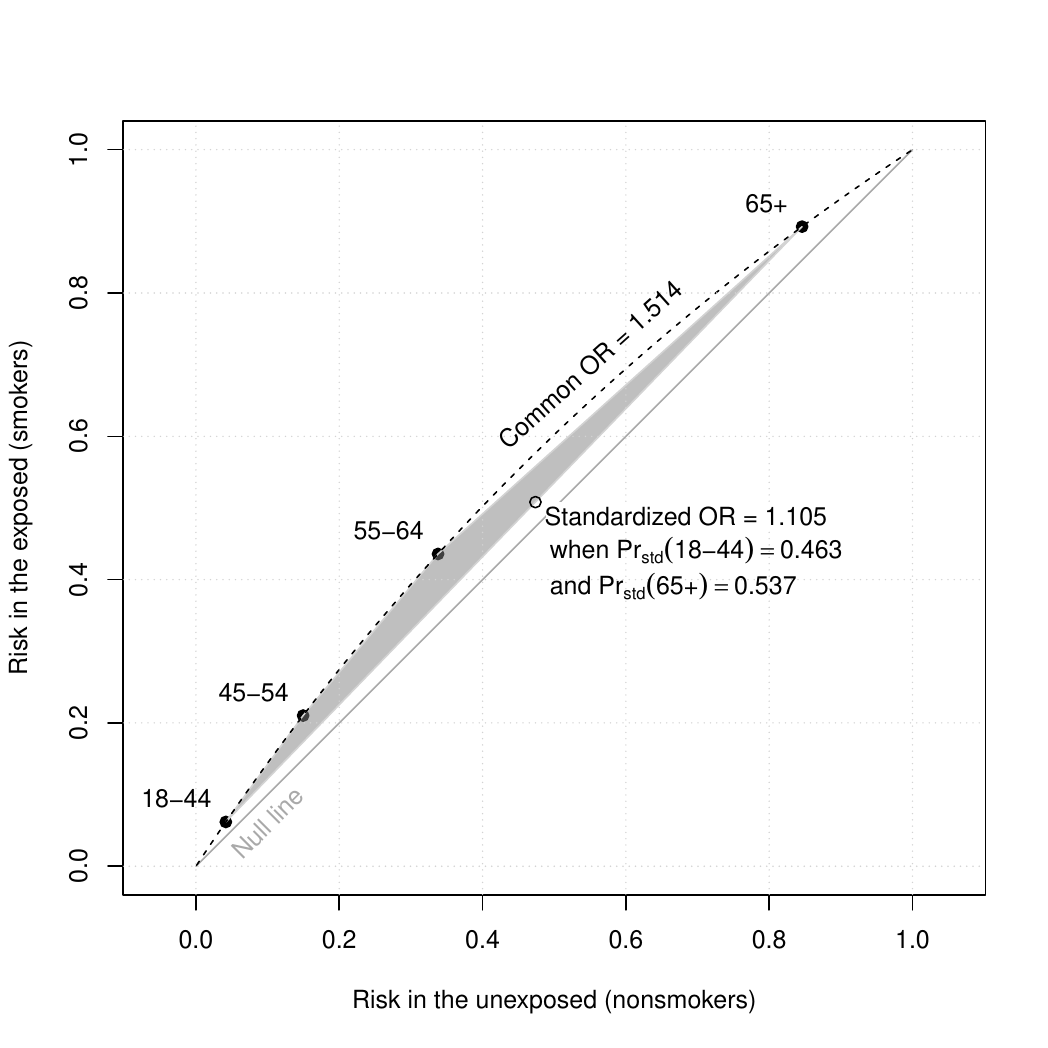}
  \caption{
    Noncollapsibility along a curved contour line for the common odds ratio with four age strata.
    Except for its vertices at the stratum-specific points (filled circles), the entire standardized hull (in gray) is between the null line and the contour for the common stratum-specific odds ratio.
    Thus, almost all standardized points have an odds ratio closer to the null.
  }
  \label{fig:noncollapsible4}
\end{figure}

When we use more than two age groups and estimate a common odds ratio or cumulative hazard ratio, the standardized hull is a polygon with vertices along the curved contour line.
This situation is illustrated in Figure~\ref{fig:noncollapsible4}, which shows the estimated stratum-specific points for four age groups from a logistic regression model with no interaction terms.
The estimated common odds ratio is $1.514$, and the likelihood ratio p-value for the null hypothesis of no age-by-smoking interaction terms is $0.81$.
The minimum possible standardized odds ratio is approximately $1.105$, which occurs in a standard population that is 46.3\% aged 18-44 years and 57.3\% aged 65+ years.

In Figure~\ref{fig:noncollapsible4}, every point in the standardized hull except its vertices has an odds ratio closer to the null than the common stratum-specific odds ratio.
Any standardized odds ratio will be closer to the null unless the standard population consists entirely of one stratum.
Conversely, a non-null stratum-specific odds ratio will move away from the null when we stratify by variable $C$ that is not a confounder or an association measure modifier.
The same is true of the cumulative hazard ratio.
Changes due to noncollapsibility will be small unless $C$ is a strong risk factor for disease (so the stratum-specific points are far apart) and the stratum-specific odds ratio or cumulative hazard ratio is far from the null (so the contour is sharply curved).
These insights would be difficult to uncover without the geometric perspective provided by Rothman diagrams.

\section{Geometry before regression}
The geometric perspective pioneered by~\citet{rothman1975pictorial} shows that association (or effect) measure modification and collapsibility are logically independent of confounding.
It shows that association measure modification should be seen as a normal occurrence, so a model with no modification on a given scale is best viewed as an approximation.
By defining collapsibility in terms of standardization rather than no confounding, we were able to see that a measure is collapsible when its contour lines are straight and that collapsibility is a property of the measure of association---one independent of the variable used for stratification.
It also provides a clear visual explanation of the consequences of collapsibility and noncollapsibility:
Standardized risk differences and risk ratios are equal to a common stratum-specific measure while standardized odds ratios and cumulative hazard ratios are almost always closer to the null.

Epidemiologists are often taught to look for confounding and causal effects in the coefficients of a regression model.
The change-in-estimates criterion for confounding is often perceived as a definition rather than a diagnostic test, which encourages the conflation of confounding and noncollapsibility.
This approach discourages the inclusion of interaction terms even though these are necessary to control confounding by an association measure modifier, and it obscures the role of model predictions and standardization in causal inference~\cite{kenah2024rothman}.
Trying to understand the relationships between confounding, association measure modification, and collapsibility through a regression model is like looking through a kaleidoscope. 
Regression models are useful tools that were never intended to teach confounding, association measure modification, or collapsibility.
As it was written over the entrance to Plato's Academy, so it can be said of regression modeling in epidemiology: ``Let no one ignorant of geometry enter here.''

\section*{Supplementary material}
The file ``GCIepi-meas.R'' contains code in \texttt{R} to produce Tables 1-2 and Figures 1-6.
Instructions are given in comments near the beginning of the file.

\section*{Acknowledgements}
I want to to thank the students of STA 6177/PHC 6937 (Applied Survival Analysis) at the University of Florida in 2014-2016 and PUBHEPI 8430 (Epidemiology 4) at Ohio State in 2019-2025 for working with early versions of this material.
I also want to thank M. Elizabeth Halloran, Wasiur R. KhudaBukhsh, Bo Lu, Nick Mandarano, Kunjal Patel, Micaela Richter, Kenneth Rothman, and Patrick Schnell for their comments.
This work was supported by National Institute of Allergy and Infectious Diseases (NIAID) grant R01 AI116770 (PI: Eben Kenah) and National Institute of General Medical Sciences (NIGMS) grant U54 GM111274 (PI: M. Elizabeth Halloran).
The content is solely the responsibility of the author and does not represent the official views or policies of NIAID, NIGMS, or the National Institutes of Health.

\printbibliography

\end{document}